# Image Encryption Using DNA Encoding, Snake Permutation and Chaotic Substitution Techniques

Waleed Ahmed Farooqui
Department of Cyber Security
PNEC NUST
Karachi, Pakistan
waleed.mscys20pnec@student.nust.edu.pk

Jawad Ahmad
Cybersecurity Center
Prince Mohammad Bin Fahd University
Alkhobar, Saudi Arabia
jahmad@pmu.edu.sa

Nadeem Kureshi
Department of MIS
PNEC NUST
Karachi, Pakistan
nadeemk@pnec.nust.edu.pk

Fawad Ahmed
Department of Cyber Security
PNEC NUST
Karachi, Pakistan
fawad@pnec.nust.edu.pk

Aizaz Ahmad Khattak
School of Computing, Engineering
and the Built Environment
Edinburgh Napier University
Edinburgh, UK
40614576@live.napier.ac.uk

Muhammad Shahbaz Khan
School of Computing, Engineering
and the Built Environment
Edinburgh Napier University
Edinburgh, UK
muhammadshahbaz.khan@napier.ac.uk

*Abstract*—Securing image data in IoT networks and other insecure information channels is a matter of critical concern. This paper presents a new image encryption scheme using DNA encoding, snake permutation and chaotic substitution techniques that ensures robust security of the image data with reduced computational overhead. The DNA encoding and snake permutation modules ensure effective scrambling of the pixels and result in efficient diffusion in the plaintext image. For the confusion part, the chaotic substitution technique is implemented, which substitutes the pixel values chosen randomly from 3 S-boxes. Extensive security analysis validate the efficacy of the image encryption algorithm proposed in this paper and results demonstrate that the encrypted images have an ideal information entropy of 7.9895 and an almost zero correlation coefficient of -0.001660. These results indicate a high degree of randomness and no correlation in the encrypted image.

## I. INTRODUCTION

The rapid increase in the development and modernization of the communication technologies had led to the vulnerability constraints in the transmission of the digital images. For the secure transmission of the images, the encryption algorithms such 3DES or AES, which are traditional encryption algorithms, can be utilized. Unlike text data, the image data is the cluster of large volumes and has color spaces, which makes it difficult for the traditional algorithms in encrypting an image. To address the limitations of the traditional encryption algorithms, innumerable techniques, based on either transform or spatial domain image processing image have been developed, with many proving their robustness in securing image from attack. As the quantum computing is becoming a reality, researchers have also started proposing and developing image encryption techniques based on quantum schemes for image data securing [1]. For the development of a robust image encryption scheme, it is necessary to include confusion and diffusion processes in it [2], [3]. Many image encryption technique propositions have been made that aims for the advancement of diffusion and confusion stage. One such example is a feature extraction based permutation technique has been proposed in [4] that scramble the pixels in the feature rick areas of the image to disrupt correlation. Similarly, a single sorund single s-box substitution technique has been proposed in [5], which proposes a lightweight substitution method for image encryption schemes.

In recent years, the use of DNA and chaos has also played an instrumental role towards the design of secure image encryption schemes [6], [7]. DNA encoding adds non-linearity and ensures effective scrambling of the pixels and prove to be an improved diffusion module for an image encryption scheme. Besides, chaos theory has also proved to be an effective source of generating pseudo-random sequences that help in generating keys and substitution indices [8], [9]. Inspired by these developments, a DNA based chaotic image encryption scheme is proposed in this paper. It is envisaged that the combination of chaos and DNA would make the proposed image encryption scheme both secure and easy to implement. A number of experiments will be carried out to gauge the effectiveness of the proposed image encryption scheme.

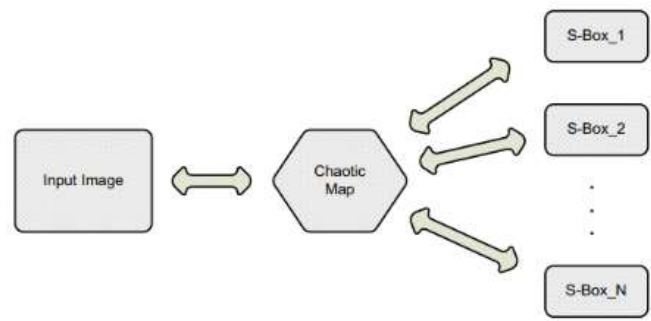

Fig. 1: Chaotic substitution using n-number of s-boxes

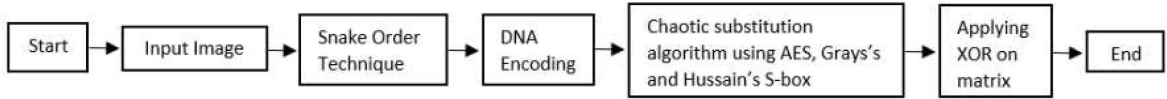

Fig. 2: Flowchart of the proposed image encryption

## II. PROBLEM STATEMENT

The substitution based image encryption schemes, such as the one proposed by Anees. et al. [10] shown in Fig. 1 offer decent encryption by utilising three S-boxes for substitution. But schemes like these are not effective in encrypting edges in an image. Only confusion based schemes are not enough in concealing the edges in an image and strong diffusion modules should be included to scramble the pixels to effectively conceal the edges and to break the correlation between the neighbouring pixels.

## III. THE PROPOSED IMAGE ENCRYPTION SCHEME

The proposed encryption scheme consists of strong confusion and diffusion modules to effectively conceal the original information including edges and breaks the correlation between the neighbouring pixels. The diffusion module consists of snake permutation and DNA encoding and the confusion module consists of chaotic substitution. The block diagram of the proposed encryption scheme is given in Fig. 2. As IoT devices have resource constraints in terms of computational power, energy, and bandwidth, the proposed algorithm has been designed to cater these constraints. The lightweight nature of the DNA encoding and Snake permutation modules ensure the efficiency of the algorithm suited for resource constrained devices. Moreover, the chaotic substitution module, in addition to being highly secure, is optimised for low latency. The proposed algorithm's focus on reduced computational complexity and low latency makes it suitable for IoT devices.

### A. Snake Permutation Technique

The snake technique is a way of rearranging the elements of a matrix in a zigzag pattern, moving back and forth across rows. The zigzag pattern is what defines the snake technique. Fig. 3 shows the working of snake technique.

In the first $4 \times 4$ matrix (left in Fig. 3), the values are in sequence from 1 to 16. In the matrix on the right, the first and third row remain the same, while the second and fourth row are reversed. Fig. 4 shows the flowchart of the snake permutation technique.

### B. DNA Encoding

In DNA encoding, digital data is stored in the form of nucelotides of DNA molecules, i.e., adenine(A), cytosine(C), guanine(G), thymine(T). By mapping binary data (0s and

**Algorithm 1** The Snake Permutation Technique
1: **Function** PermSnake($Perm_s$)
2: $[M, N] \leftarrow$ size($Perm_s$)
3: $Perm_{\text{snake}} \leftarrow$ Array initialization having size $M \times N$
4: **for** $i = 1$ to $M$ **do**
5:    **if** value of $i$ is **odd then**
6:       $Perm_{\text{snake}}[i, :] \leftarrow Perm_s[i, :]$ {L-R}
7:    **else**
8:       $Perm_{\text{snake}}[i, :] \leftarrow$ Inverse($Perm_s[i, :]$) {R-L}
9:    **end if**
10: **end for**
11: **ret** $Perm_{\text{snake}}$
12: **End**

**Algorithm 2** DNA Encoding
1: **Function** DNAEnc($Y$)
2: $[M, N] \leftarrow$ size($Y$)
3: $A \leftarrow$ Initialisation of an array having size of $M \times N$
4: **for** $i = 1$ to $M$ **do**
5:    **for** $j = 1$ to $N$ **do**
6:       $x \leftarrow$ Converting $Y$ to binary (8 bit)
7:       $A \leftarrow$ Mapping bits in $x$ to DNA nucleotides (A, C, G, T)
8:    **end for**
9: **end for**
10: **ret** $A$
11: **End**

1s) to these nucleotides, vast amounts of information can be compactly stored in a biologically stable medium. In our scheme, we have converted the matrix pixels first into 8 bit binary form, then separated it 8 bits into two parts each as seen in Fig. 6.

The binary sequences used in the code are shown in Table I. Keeping in view the encoding in Table I, 11000110 becomes 'ATGC'. Fig. 6 shows the flow of DNA Encoding as used in our proposed scheme.

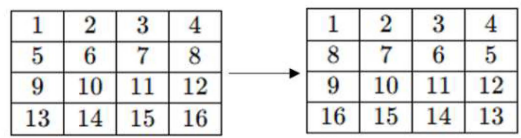

Fig. 3: Snake Technique

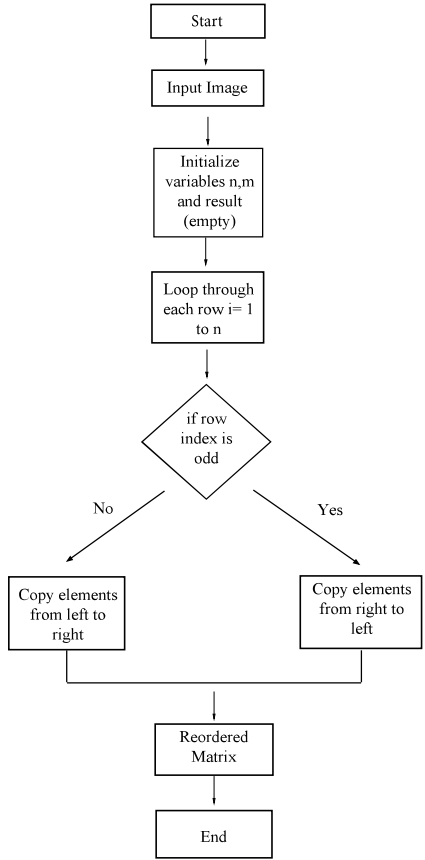

Fig. 4: The snake permutation technique

11000110 ⟶ '11' '00' '01' '10'

Fig. 5: Mapping to the nucleotides

## C. Steps for Proposed Encryption Algorithm

**Step 1:** First, we take a plain image of 256 x 256 pixels. In our case, we have used cameraman and Baboon image for testing.

**Step 2:** After reading the matrix, the plaintext image undergoes permutation using the Snake technique. This Snake technique as described in Section III-A scrambles the pixels extensively, making them unrecognizable compared to the original image.

**Step 3:** The third step applies DNA Encoding to the permuted matrix from the Snake technique. In a 256 x 256

TABLE I: Binary Sequences

| Nucleotide | Binary Sequence |
|---|---|
| A | 11 |
| T | 00 |
| C | 10 |
| G | 01 |

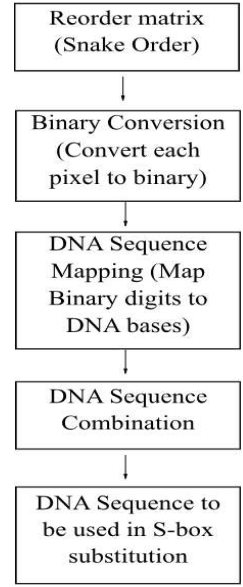

Fig. 6: DNA Encoding flowchart as used in the proposed scheme

image, each pixel value is converted to binary and divided into four 2-bit segments. For example, a pixel value of 56 becomes 00111000, split into "00", "11", "10", and "00". Each segment is mapped to nucleotides A, T, G, or C, based on predefined sequences shown in Fig. 5. This generates a DNA sequence. The generated DNA sequence will have 262,144 characters, formatted as ACGTCGTCATN ,,,,,N. It will then undergo Random Permutation to enhance its randomness and complexity.

**Step 4:** The next step converts the DNA sequence back into matrix format. Each nucleotide—A, C, G, or T—is mapped to a predefined binary value (as shown in TABLE 1): A as 00, C as 01, G as 10, and T as 11. For example, the sequence AGCT translates to the binary sequence 00100111.

A main loop will iterate 65,536 times, corresponding to the 256 x 256 matrix size. In each iteration, sets of 4 binary values will be combined to reconstruct the matrix. For example, the sequence ACGT will loop through 4 iterations to form an 8-bit binary value. After completing the iterations, the result will be reshaped into the 256 x 256 matrix format.

**Step 5:** In this step, a chaotic sequence for the selection of S-boxes is generated using the logistic map, whic is defined as:

$$x_{n+1} = r \cdot x_n \cdot (1 - x_n)$$

This map generates a matrix with integer values strictly between 0, 1, and 2, which correspond to S-box 1, S-box 2,

and S-box 3, respectively.

**Step 6:** In this step, we process the matrix from DNA encoding by iterating 65,536 times, corresponding to the matrix size of $256 \times 256$. Each value is converted to binary. For instance, a value of 85 becomes 01010101. This binary value is split into two 4-bit segments, i.e., 0101 and 0101, which converts to decimal values of 5 and 5. Adding 1 to these values gives the index of the value to selected from the S-box, e.g., adding 1 to $5 \times 5$ gives $6 \times 6$. The value at the index $6 \times 6$ in the S-box is chosen to substitute the original pixel value in the image. This process creates a final matrix combining values from all three S-boxes.

**Step 7:** In the final step, we perform an XOR operation between the resultant matrices. This operation will yield the cipher image.

## IV. STATISTICAL SECURITY ANALYSIS

We conducted various statistical security analyses including correlation, entropy, homogeneity, contrast, and energy metrics. The histogram analysis has also been performed. All tests were performed using the Baboon image and Cameraman image.

### A. Correlation Analysis and Comparison

Correlation measures how closely the neighbouring pixels are related to each other. Natural images usually show maximum or high correlation due to similar values in neighboring pixels. A good encryption algorithm, however, disrupts this similarity between the neighbouring pixels and correlation of the encrypted image should be as lowest as possible, ideally zero. Correlation is calculated using the following formula.

$$r = \frac{\sum_{x=1}^{n}(M_x - \bar{M})(N_x - \bar{N})}{\sqrt{\sum_{x=1}^{n}(M_x - \bar{M})^2 \sum_{x=1}^{n}(N_x - \bar{N})^2}}$$

Here:
- $\bar{M}$ and $\bar{N}$ correspond to the mean of the variables $M$ and $N$.
- $n$ represents the total number of data points.

The Correlation values when compared to the other schemes are given Table II.

TABLE II: Correlation values of cipher images

| Source | Correlation |
|---|---|
| Hussain et al. [11] | 0.0336 |
| Anees et al. [10] | 0.0201 |
| Jan Sher et al. [12] | -0.0012 |
| **The Proposed Scheme** | **-0.0016603** |

Fig. 7 displays the correlation coefficients of the baboon image. It is evident that all three coefficients, i.e., vertical, horizontal, and diagonal correlation coefficients are dispersed and have effectively broken the correlation between the neighbouring pixels.

### B. Entropy Analysis and Comparison

Entropy measures the randomness in an image, reflecting how unpredictable pixel values are after encryption. High entropy indicates effective encryption with uniform pixel value distribution, reducing exploitable patterns. It is calculated based on the probability distribution of pixel intensities. Mathematically, entropy $H$ of an image can be expressed as:

$$H(X) = -\sum_{k=1}^{n} x(n_k) \log_2 x(n_k)$$

where $X$ represents the set of pixel values,
$x(n_k)$ is the probability of occurrence of the pixel value $n_k$,
$n$ is the number of distinct pixel values.

The results of entropy including comparison with other schemes are given in the Table III. A perfectly encrypted image should exhibit an entropy value close to the theoretical maximum, suggesting that the pixel values are equally likely and independent of each other.

TABLE III: Information entropy values of cipher images

| Source | Entropy |
|---|---|
| Hussain et al. [11] | 7.8815 |
| Anees et al. [10] | 7.8523 |
| Jan Sher et al. [12] | 7.9884 |
| **The Proposed Scheme** | **7.9895** |

### C. Homogeneity Analysis and Comparison

Homogeneity measures how similar or uniform the adjacent pixels of an image are. In an encrypted image, high homogeneity indicates uniformly distributed pixel values, reducing discernible patterns and enhancing security against statistical attacks. The formula is given as: -

$$H = \sum_{x,y} \frac{P(x,y)}{1 + |x-y|}$$

- $P(x,y)$ represents the co-occurrence matrix, where each element $P(x,y)$ is the joint probability of pixel pairs with intensities $x$ and $y$.
- $|x-y|$ corresponds to the absolute difference between the two pixel values $x$ and $y$.

The homogeneity index $H$ gauges how closely the co-occurrence matrix distribution aligns with its diagonal. A higher value indicates more uniform pixel values, which is typically undesirable in image encryption as it suggests lower randomness. The results of Homogeneity are given in Table IV.

TABLE IV: Homogeneity values of several cipher images obtained by different algorithms

| Source | Homogeneity |
|---|---|
| Hussain et al. [11] | 0.3452 |
| Anees et al. [10] | 0.4208 |
| Jan Sher et al. [12] | 0.3973 |
| **The Proposed Scheme** | **0.39233** |

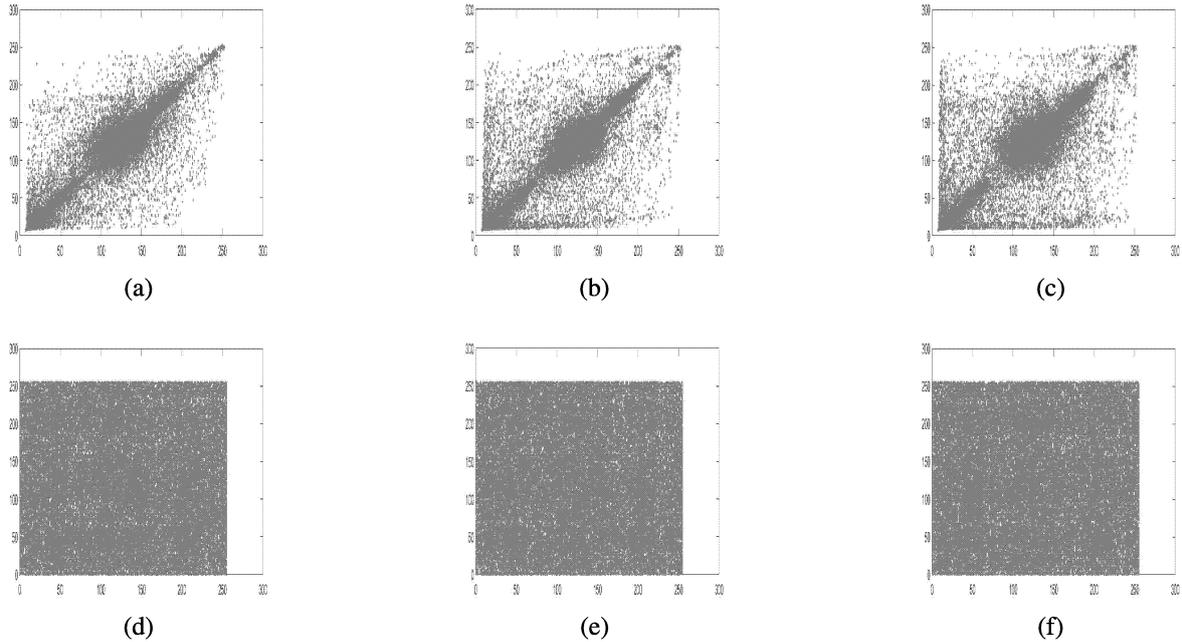

Fig. 7. Correlation Analysis of Cameraman Image: (a), (b) and (c) represents Vertical, Horizontal and Diagonal Coefficients of an Plain Image. (d), (e) and (f) represents Vertical, Horizontal and Diagonal Coefficients of an Cipher Image

*D. Contrast Analysis and Comparison*

The contrast refers to the difference in the pixel intensities of two neighbouring or adjacent pixels. Mathematically, contrast can be quantified using:

$$C = \sum_{x,y}(x - y)^2 P(x,y)$$

Here, $P(x,y)$ is the co-occurrence matrix element representing the joint probability of pixel pairs with intensities $x$ and $y$.

The results of contrasts are given in Table V. Higher contrast signifies greater disparity between pixel values, reflecting a more effective encryption process. The results indicate a higher value of the contrast for the cipher image.

TABLE V: Contrast values of cipher images

| Source | Contrast |
| --- | --- |
| Hussain et al. [11] | 6.8815 |
| Anees et al. [10] | 8.3301 |
| Jan Sher et al. [12] | 9.9797 |
| **The Proposed Scheme** | **10.1515** |

*E. Energy Analysis and Comparison*

Energy measures the sum of squared elements in the co-occurrence matrix, reflecting pixel intensity uniformity. A low energy value in an cipher image indicates even distribution of pixel values with no dominant patterns, enhancing security by minimizing identifiable features. Mathematically, energy $E$ can be quantified using the following formula derived from the co-occurrence matrix:

$$E = \sum_{i,j}[P(i,j)]^2$$

where:

- $P(i,j)$ represents the co-occurrence matrix, where each element $P(i,j)$ is the joint probability of pixel pairs with intensities $i$ and $j$.

The Table VI shows the results that indicates a low value of energy for the tests images encrypted by the proposed algorithm, validating the efficacy of the proposed scheme.

TABLE VI: Energy values of cipher images

| Source | Energy |
| --- | --- |
| Hussain et al. [11] | 0.0128 |
| Anees et al. [10] | 0.0176 |
| Jan Sher et al. [12] | 0.0159 |
| **The Proposed Scheme** | **0.015865** |

*F. Histogram Analysis*

Histogram analysis compares the pixel intensity distributions of the Plain and Cipher images. An securely encrypted image should display a histogram having uniform distribution of pixel intensities. The Plain image of Baboon Fig. 8(a) and Cameraman Fig. 8(e) and their histogram as shown in Fig. 8(c) and 8(g) respectively typically shows peaks and troughs corresponding to the image content's pixel value distribution. On the other hand, a uniform histogram of Baboon Fig. 8(d) and Cameraman Fig. 8(h) in the cipher images of Baboon Fig. 8(b) and Cameraman Fig. 8(f) suggests that the encryption algorithm effectively masks the Plain images's features. This uniformity means that each pixel value is equally probable, complicating any attempt by an attacker to infer information about the Plain images.

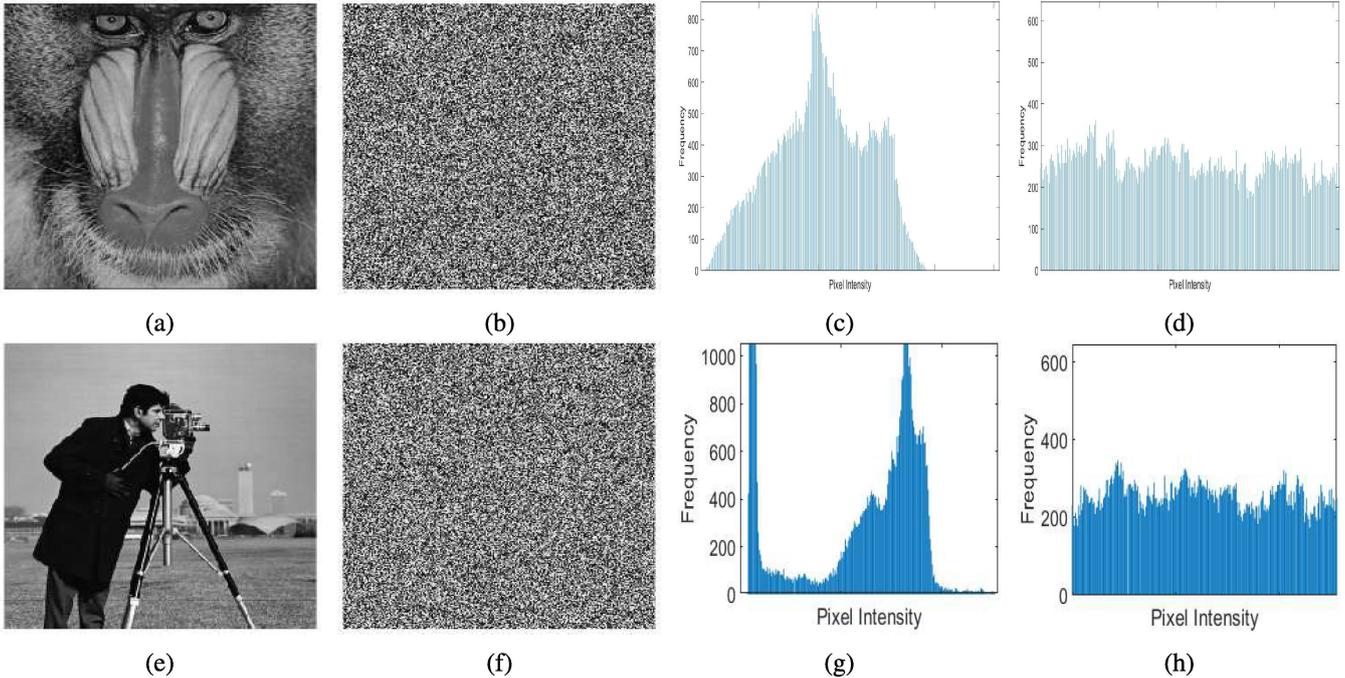

Fig. 8. Histogram Analysis of Plain Image and Cipher Image, Baboon(a)-(d), Cameraman (e)-(h)

## V. Conclusion

In this paper, a robust image encryption algorithm was proposed and evaluated, which integrated snake permutation, DNA encoding and chaotic maps based substitution. The proposed scheme was tested on various image datasets, and the results consistently showed that the proposed method achieves robust encryption with minimal computational overhead. The algorithm proposed in this paper utilised chaotic map to generate highly random and non-linear sequences to be used as substitution and permutation indices. The successful reduction in pixel correlation and an ideal information entropy indicates that the algorithm proposed in this paper is highly effective in encrypting images securely. Future work could explore further optimizations and extensions of this method, including its application to video encryption and real-time communication systems.

## References

[1] M. S. Khan, J. Ahmad, A. Al-Dubai, N. Pitropakis, B. Ghaleb, A. Ullah, M. A. Khan, and W. J. Buchanan, "Chaotic quantum encryption to secure image data in post quantum consumer technology," *IEEE Transactions on Consumer Electronics*, pp. 1–1, 2024.

[2] C. Shannon, "Claude shannon," *Information Theory*, vol. 3, p. 224, 1948.

[3] A. Afifi, "A chaotic confusion-diffusion image encryption based on henon map," *International Journal of Network Security & Applications (IJNSA) Vol*, vol. 11, 2019.

[4] M. S. Khan, J. Ahmad, A. Al-Dubai, Z. Jaroucheh, N. Pitropakis, and W. J. Buchanan, "Permutex: Feature-extraction-based permutation — a new diffusion scheme for image encryption algorithms," in *2023 IEEE 28th International Workshop on Computer Aided Modeling and Design of Communication Links and Networks (CAMAD)*, pp. 188–193, 2023.

[5] M. S. Khan, J. Ahmad, H. Ali, N. Pitropakis, A. Al-Dubai, B. Ghaleb, and W. J. Buchanan, "Srss: A new chaos-based single-round single s-box image encryption scheme for highly auto-correlated data," in *2023 International Conference on Engineering and Emerging Technologies (ICEET)*, pp. 1–6, 2023.

[6] J. Zhao, S. Wang, and L. Zhang, "Block image encryption algorithm based on novel chaos and dna encoding," *Information*, vol. 14, no. 3, p. 150, 2023.

[7] D. Ravichandran, A. Banu S, B. Murthy, V. Balasubramanian, S. Fathima, and R. Amirtharajan, "An efficient medical image encryption using hybrid dna computing and chaos in transform domain," *Medical & biological engineering & computing*, vol. 59, pp. 589–605, 2021.

[8] Z.-H. Guan, F. Huang, and W. Guan, "Chaos-based image encryption algorithm," *Physics letters A*, vol. 346, no. 1-3, pp. 153–157, 2005.

[9] B. Zhang and L. Liu, "Chaos-based image encryption: Review, application, and challenges," *Mathematics*, vol. 11, no. 11, p. 2585, 2023.

[10] A. Anees, A. M. Siddiqui, and F. Ahmed, "Chaotic substitution for highly autocorrelated data in encryption algorithm," *Communications in Nonlinear Science and Numerical Simulation*, vol. 19, no. 9, pp. 3106–3118, 2014.

[11] I. Hussain, A. Anees, and A. Algarni, "A novel algorithm for thermal image encryption," *Journal of Integrative Neuroscience*, vol. 17, pp. 447–461, Sep 2018.

[12] J. S. Khan and J. Ahmad, "Chaos based efficient selective image encryption," *Multidimensional Systems and Signal Processing*, vol. 30, no. 2, pp. 943–961, 2019.